\documentclass[10pt,aps,twocolumn,showpacs,superscriptaddress,prl]{revtex4-1}
\usepackage{graphicx}
\usepackage{amsmath}
\usepackage{amssymb}
\usepackage[sort&compress]{natbib}
\usepackage[latin1]{inputenc}
\usepackage{tgtermes}
\usepackage{pdfpages}

\makeatletter
\AtBeginDocument{\let\LS@rot\@undefined}
\makeatother

\newcommand{\uu}[1]{{\boldsymbol #1}}

\def\xb{\uu{x}}

\def\nb{\uu{n}}

\def\etab{\uu{\eta}}

\begin{document}

\title{Spatiotemporal Self-Organization of Fluctuating Bacterial Colonies} \author{Tobias
  \surname{Grafke}} \affiliation{Courant Institute, New York
  University, 251 Mercer Street, New York, NY 10012, USA}
\author{Michael E. \surname{Cates}} \affiliation{DAMTP, Centre for
  Mathematical Sciences, Wilberforce Road, Cambridge, CB3 0WA, United
  Kingdom} \author{Eric \surname{Vanden-Eijnden}} \affiliation{Courant
  Institute, New York University, 251 Mercer Street, New York, NY
  10012, USA}

\date{\today}

\begin{abstract}
  We model an enclosed system of bacteria, whose motility-induced
  phase separation is coupled to slow population dynamics.  Without
  noise, the system shows both static phase separation and a limit
  cycle, in which a rising global population causes a dense bacterial
  colony to form, which then declines by local cell death, before
  dispersing to re-initiate the cycle. Adding fluctuations, we find
  that static colonies are now metastable, moving between spatial
  locations via rare and strongly nonequilibrium pathways, whereas the
  limit cycle becomes almost periodic such that after each
  redispersion event the next colony forms in a random location.
  These results, which hint at some aspects of the biofilm-planktonic
  life cycle, can be explained by combining tools from large deviation
  theory with a bifurcation analysis in which the global population
  density plays the role of control parameter.
\end{abstract}

% PACS
% 87.10.Mn Biophysics: Stochastic modeling
% 87.17.Jj Biophysics: Cell locomotion, chemotaxis
% 05.40.-a General: Fluctuation phenomena

\maketitle

Because they are not bound by the standard laws of equilibrium
thermodynamics, active materials such as bird flocks
\cite{ballerini-cabibbo-candelier-etal:2008}, motile bacteria
\cite{marchetti-joanny-ramaswamy-etal:2013}, self-organizing
bio-polymers \cite{schaller-weber-semmrich-etal:2010,
  sumino-nagai-shitaka-etal:2012}, or man-made self-propelled
particles \cite{howse-jones-ryan-etal:2007} have many more routes
towards self-assembly and self-organization than systems whose
dynamics satisfy detailed-balance. Motility-induced phase separation
(MIPS) is one example
\cite{wittkowski-tiribocchi-stenhammar-etal:2014,
  cates-tailleur:2015}. MIPS arises naturally in systems of
self-propelled particles whose locomotive speed decreases
monotonically with density, through a feedback in which particles
accumulate where they move slowly and vice-versa.  This provides a
generic path to pattern formation in motile agents, including both
micro-organisms such as \emph{E. coli}~\cite{shapiro:1995,
  tailleur-cates:2008}, where slowdown can be caused by quorum
sensing~\cite{parsek-greenberg:2005, liu-fu-liu-etal:2011,
  fu-tang-liu-etal:2012, solano-echeverz-lasa:2014}, and synthetic
colloidal analogues where slowdown is caused by crowding
\cite{fily-marchetti:2012, palacci-sacanna-steinberg-etal:2013,
  buttinoni-bialke-kuemmel-etal:2013, cates-tailleur:2013,
  cates-tailleur:2015}. The simplest theories of MIPS describe phase
separation through an effective free energy functional
\cite{cates-tailleur:2015}, although active gradient terms can alter
this structure and the resulting coexistence
behavior~\cite{wittkowski-tiribocchi-stenhammar-etal:2014,
  cates-tailleur:2015}.
  
MIPS, can be arrested by the birth and death of particles. The
simplest (logistic) population dynamics has for uniform systems a
stable fixed point at some carrying capacity $\rho_0$, with decay
towards this from higher or lower densities. If $\rho_0$ lies within
the miscibility gap of a phase separation, the uniform state at
$\rho_0$ is unstable, but so is a state of coexisting bulk phases
(each of which would have nonstationary density). This can result in
stationary patterns of finite wavelength, in which bacteria reproduce
in dilute regions, migrate by diffusive motility to dense ones, become
immotile, and there die off by overcrowding
\cite{cates-marenduzzo-pagonabarraga-etal:2010,cates-tailleur:2015}.
In~\cite{cates-marenduzzo-pagonabarraga-etal:2010} a field theoretic
description of such arrested phase separation was proposed to capture
the variety of bacterial colony patterns earlier observed in
experiments \cite{murray:2003}.

The work of~\cite{cates-marenduzzo-pagonabarraga-etal:2010} mainly
describes stationary spatial rather than spatiotemporal patterns, and
crucially neglects the effects of intrinsic fluctuations in population
density and motility of the bacteria. These fluctuations are always
present in finite systems and they are relevant to the nature and
stability of the self-assembling structures. One purpose of the
present Letter is to explore these effects in situations where birth
and death processes are slow compared to diffusive exploration
times. Another is to show that entirely different physics from that
reported in~\cite{cates-marenduzzo-pagonabarraga-etal:2010} can arise,
involving spatiotemporal rather than static patterning. For simplicity
we focus on one-dimensional systems whose extent is smaller than the
natural length scale for the spacing between dense patches, controlled
by the distance a particle can move during its lifetime. Accordingly,
only one dense bacterial domain of low motility (hereafter `colony')
is present at a time, with the remaining bacteria in a dilute and
motile (i.e., `planktonic') phase. These simplifications allow us to
focus on temporal aspects of the pattern formation, and also to give a
thorough analysis of the role of noise terms. Our qualitative
conclusions, however, apply in higher dimensions.

Concretely, we identify two regimes in which the fluctuations induce
nontrivial self-organization pathways. The first arises in situations
where the deterministic dynamics possess time-periodic solutions
driven by the interplay between MIPS and the competition for
resources: in the resulting limit cycles a colony of bacteria
periodically appears and disappears at a fixed location in the
domain. In this regime, the fluctuations, no matter how small, are
shown to have a drastic impact: they allow the colony to explore its
environment by randomly jumping from one location to another each time
the system revisits its spatially homogeneous planktonic state. With
periodic spatial boundary conditions the colony now appears and
disappears at spatially random locations; if boundary conditions
instead favor localization at container walls, a random choice of wall
is made each cycle.  It is striking that this behavior, which
resembles the biofilm-planktonic lifecycle of several
micro-organisms~\cite{kostakioti-hadjifrangiskou-hultgren:2013} can
arise from the combination of MIPS and logistic growth alone --
particularly as the bacterial quorum sensing response, a likely cause
of MIPS
\cite{tailleur-cates:2008,cates-marenduzzo-pagonabarraga-etal:2010} is
now thought also to be directly implicated in biofilm
dispersion~\cite{solano-echeverz-lasa:2014}. We are not suggesting
that the combination of MIPS and logistic growth directly explains the
complex life cycle of real biofilms, but it may nonetheless be a
contributory factor in its evolution from simpler beginnings.

In the second regime, we show that the deterministic theory predicts a
single static colony with multiple stable locations. In these
situations the intrinsic fluctuations induce rare, noise-activated
transitions via the uniform state from one such metastable pattern to
another.  (This again resembles the biofilm lifecycle, but now with
random intervals between dispersion events.)  As shown below, the rate
and mechanism of these transitions can be completely characterized
using tools from large deviation theory
(LDT)~\cite{freidlin-wentzell:2012}.  The transition pathways are
intrinsically out-of-equilibrium and involve a subtle interplay
between phase-separation and reproduction. In particular these
transitions do not follow the deterministic relaxation path in
reverse, as one would predict from standard equilibrium arguments in
systems with microscopic reversibility.

\begin{figure*}[tb]
  \begin{center}
    \hfill
    \includegraphics[width=0.3\linewidth]{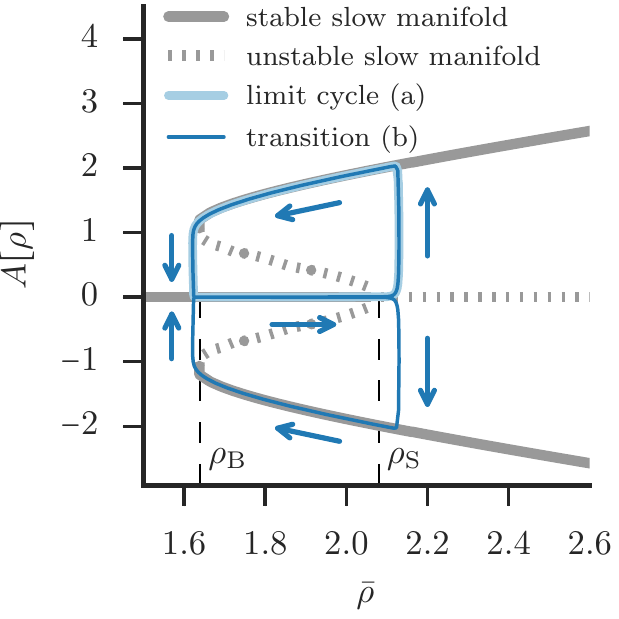}
    \hfill
    \includegraphics[width=0.6\linewidth]{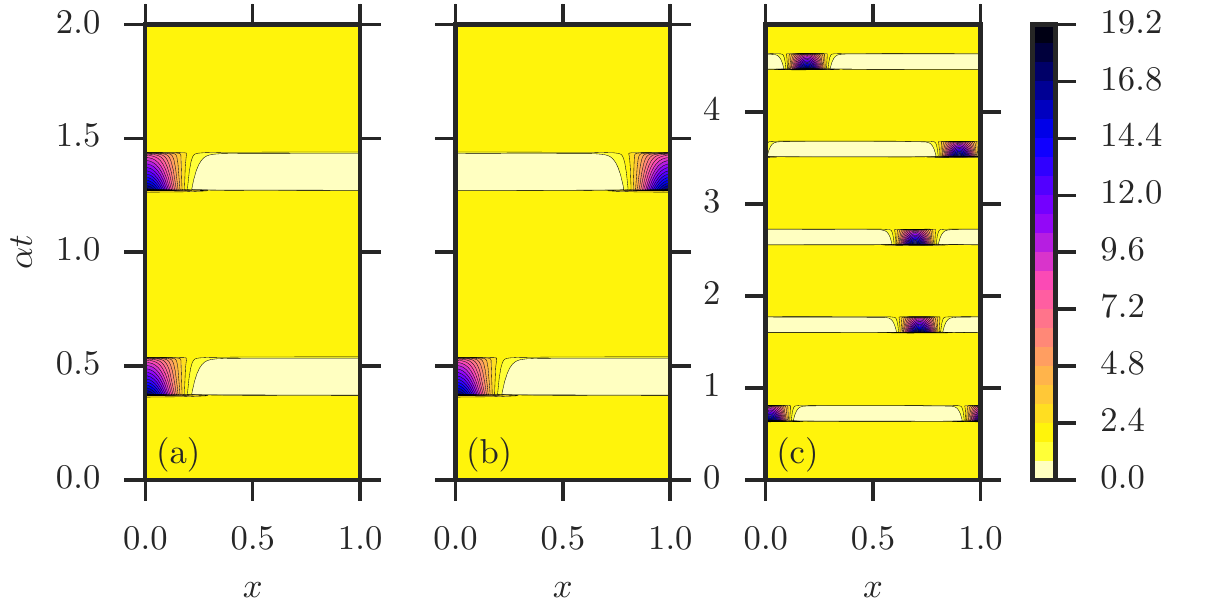}
    \hfill
  \end{center}
  \caption{\emph{Left panel:} Bifurcation diagram in a 2-dimensional
    projection. The $x$-axis shows the spatial mean of $\rho$, the
    $y$-axis shows its signed asymmetry $A[\rho]$ (see text for
    details). For carrying capacities
    \mbox{$\rho_{\text{S}}<\rho_0<\rho_0^c$}, no stable fixed point of the
    dynamics exist. A limit cycle (light blue) and a trajectory with a
    small amount of noise (dark blue) that exhibits a transition to
    the lower stable branch are projected into the
    diagram. \emph{Right panels (a), (b), and (c)}: The time-evolution
    of the limit cycle with a colony located at the left boundary
    (upper cycle on the left panel) is shown in subplot (a).
    Fluctuations allow the solution to randomly jump between the limit
    cycles with colonies located at the left or right of the domain,
    as shown in subplot (b). Subplot (c) depicts the almost periodic
    regime in a spatially periodic domain, where the fluctuations
    randomize the colony location. \label{fig:projection1}}
\end{figure*}
Following~\cite{cates-marenduzzo-pagonabarraga-etal:2010}, we focus on
motile bacteria whose motile diffusivity
\mbox{$D(\rho)=\tfrac12 v^2(\rho)\tau/d$}~\cite{tailleur-cates:2008} depends on
their local density~$\rho$ via the swim speed $v(\rho)$. Here $\tau$
is a rotational relaxation time and $d$ is dimensionality.  Because
the swim speed varies in space through $\rho$, the particles have a
mean drift velocity \mbox{$\boldsymbol{V}(\rho) =
  -\tfrac12D'(\rho)\nabla \rho$}~\cite{tailleur-cates:2008}.
Neglecting fluctuations, these effects can be combined into the
following equation governing the evolution of the bacterial
density~$\rho(\xb,t)$:
\begin{equation}
  \label{eq:model0}
  \partial_t \rho = \nabla \cdot (\mathcal{D}_e(\rho) \nabla \rho)
  -\delta^2 \nabla\cdot \left( \rho D(\rho) \nabla \Delta \rho\right)\,.
\end{equation}
In this equation the diffusivity and drift have been combined into a
collective diffusivity
$\mathcal{D}_e(\rho)=D(\rho)+\tfrac12\rho\,D'(\rho)$, and a
higher-order gradient, or regularizing, term proportional to
$\delta^2$ has been added to account for the fact that the bacteria
only sense each other's influence over a finite distance
$\delta>0~$\footnote{In general this effect leads to a nonlocal
  equation, out of which~\eqref{eq:model0} emerges at leading order
  after expansion in
  $\delta\ll1$~\cite{grafke-hirschberg-vanden-eijnden:2017}. }.

Our choice of regularizer is the simplest form to emerge from explicit
coarse-graining of the microscopic dynamics of active particles
\cite{tailleur-cates:2008, speck-bialke-menzel-etal:2014,
  cates-tailleur:2015}. With this form, and in the absence of
birth-death processes, the physics of MIPS, including all noise
contributions, maps exactly onto an equilibrium model of phase
separation. The regularizer then derives from a square gradient
contribution in an underlying free energy that describes passive
Brownian particles with attractive
interactions~\cite{cates-tailleur:2015}. (A more detailed
coarse-graining gives further gradient terms that are not
representable by any free energy; these shift the MIPS phase
boundaries only slightly~\cite{cates-tailleur:2015,
  stenhammar-tiribocchi-allen-etal:2013}.) Thus our model, whose
microscopic derivation is in the supplemental
material~\setcounter{footnote}{136}\footnote{See supplemental material
  at [URL will be inserted by publisher] for a derivation from a
  microscopic model}, includes a fluctuating noise term in
Eq.~\eqref{eq:model0} such that its dynamics obey detailed balance
(DB) with the free energy
\begin{equation}
  \label{eq:energy1}
    E[\rho] = \int_\Omega\left(\rho\log\rho - \rho + f(\rho)
      +\tfrac12 \delta^2 |\nabla \rho|^2\right)d\xb  \,.
\end{equation}
Eq.~\eqref{eq:model0} then takes the form of a generalized gradient
flow, \mbox{$\partial_t \rho = -M(\rho) (\delta F/\delta \rho)$}, with
nonlinear mobility operator $M(\rho)\xi=\nabla\cdot(\rho
D(\rho)\nabla\xi)$~\cite{mielke-renger-peletier:2016}. The DB property
of Eq.~\eqref{eq:model0} is inessential since it is violated by the
birth and death terms added below
\setcounter{footnote}{1336}\footnote{Detailed balance violation arises
  because the rates for birth and death are unaffected by the motility
  induced ``attractions'' (whose true origin is kinetic) whereas
  genuine enthalpic attractions would require the birth and death
  rates to become explicitly density dependent
  \cite{lefever-carati-hassani:1995,
    glotzer-di_marzio-muthukumar:1995, glotzer-stauffer-jan:1994,
    glotzer-stauffer-jan:1995}.}  but it simplifies the analysis when
these are small. The regularizer used here is an improvement on the
one adopted in~\cite{cates-marenduzzo-pagonabarraga-etal:2010}, which
is purely phenomenological and does not emerge from any known
coarse-graining of an active-particle model.

%Our choice of regularizer, which differs from that of
%\cite{cates-marenduzzo-pagonabarraga-etal:2010}, allows
%\eqref{eq:model0} to be cast as an effective equilibrium model.
%%
%That is,
%when a fluctuating noise term is
%added to \eqref{eq:model0},
%the resulting equation is in detailed balance (DB) with the free energy 
%\begin{equation}
%  \label{eq:energy1}
%    F[\rho] = \int_\Omega\left(\rho\log\rho - \rho + f(\rho)
%      +\tfrac12 \delta^2 |\nabla \rho|^2\right)d\xb  \,,
%\end{equation}
%where $f'(\rho) = \log\,D(\rho)/2$, even though the underlying
%microscopic dynamics are not. Eq.~\eqref{eq:model0} then takes the
%form of a generalized gradient flow, \mbox{$\partial_t \rho = -M(\rho)
%(\delta F/\delta \rho)$}, with nonlinear mobility operator
%$M(\rho)\xi=\nabla\cdot(\rho
%D(\rho)\nabla\xi)$~\cite{mielke-renger-peletier:2016}.  The DB
%property is destroyed by the birth and death terms addressed below,
%but simplifies the analysis when these are small.

Crucially, whenever $d\ln v/d\ln\rho<-1$, we have
$\mathcal{D}_e(\rho)<0$ in \eqref{eq:model0}. This is the spinodal
regime of local instability for MIPS~\cite{tailleur-cates:2008}. We
assume $v(\rho) = v_0 e^{-\lambda\rho/2}$ where $v_0$ is the speed of
an isolated particle and
$\lambda>0$~\cite{cates-marenduzzo-pagonabarraga-etal:2010}. In a
$d$-dimensional box \mbox{$\Omega=[0,L]^d$}, after
non-dimensionalization via $\tau v_0^2 = \lambda = 1$,
\eqref{eq:model0} reduces to
\begin{equation}
  \label{eq:model}
  \begin{aligned}
    \partial_t \rho &=
    \nabla\cdot\big((1-\tfrac12\rho)e^{-\rho}\nabla\rho-\delta^2\rho
    e^{-\rho}\Delta\nabla\rho\big)\,.
  \end{aligned}
\end{equation}
We study~\eqref{eq:model} for $\xb\in\Omega=[0,1]^d$ with, unless
otherwise stated, Neumann boundary conditions: $\hat \nb
\cdot\nabla\rho=0\,\forall\,\xb\in\partial\Omega$ for $\hat \nb$
normal to the boundary. These localize the dense phase at a wall, to
minimize its interfacial energy.

The mean bacterial density, $\bar\rho=|\Omega|^{-1}\int_\Omega
\rho(x)\,d\xb$, is conserved by \eqref{eq:model} and controls the
phase separation. Fig.~\ref{fig:projection1} (left) shows the
bifurcation diagram obtained when \mbox{$d=1$} and
\mbox{$\delta^2=2\cdot 10^{-3}$} by projecting the fixed points
of~\eqref{eq:model} onto the $(\bar\rho,A)$ plane, where the `signed
asymmetry' is defined as $A[\rho]=\smash{\int_0^{1/2}\rho(x)\,dx} -
\smash{\int_{1/2}^1 \rho(x)\,dx}$.  A linear stability analysis finds
the spatially uniform solution to be stable if
$\bar\rho<\rho_{\text{S}} = 2/(1-2\delta^2\pi^2)$. (This becomes the
bulk spinodal condition in the large system, or small $\delta$,
limit.)  A subcritical pitchfork bifurcation occurs at
$\bar\rho=\rho_{\text{S}}$, where two unstable and one stable branches
merge into a single unstable branch. The remaining two stable
branches, $\rho_\text{L}(x)$ and $\rho_\text{R}(x)$, correspond to
colony formation on either the left or the right wall. Once present,
each such phase-separated state remains stable down to a `binodal'
density $\bar\rho=\rho_{\text{B}}<\rho_{\text{S}}$~\footnote{For the
  particular choice in \eqref{eq:model}, $\rho_{\text{B}}\to 0$ in the
  large system limit but for a finite system the lower stability limit
  $\rho_{\text{B}}<\rho_{\text{S}}$ remains nonzero, albeit now
  depending on $\bar\rho$ as well as $\delta$.}, whereas for
$\bar\rho< \rho_{\text{B}}$ the colony redisperses diffusively.

\begin{figure}[tb]
  \begin{center}
    \includegraphics[width=0.9\columnwidth]{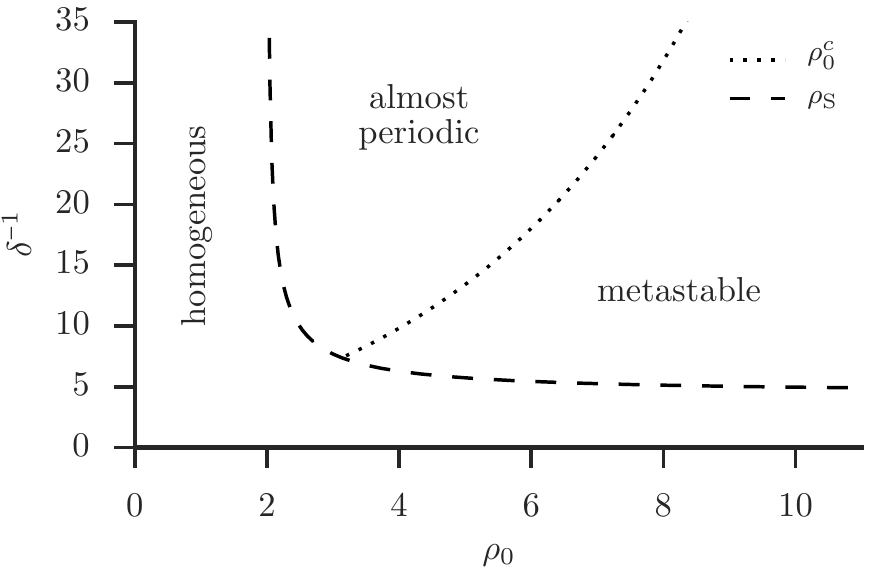}
  \end{center}
  \caption{Phase diagram a function of $\rho_0$ and $\delta$, showing
    regimes with homogeneous, almost periodic, and metastable
    solutions. The dashed line shows
    $\rho_{\text{S}} = 2/(1-2\delta^2\pi^2)$, the dotted line
    $\rho_0^c$.\label{fig:phasediag}}
\end{figure}
These transitions in static stability become {\em dynamically}
significant once the mean bacterial density $\bar\rho$ is allowed to
change by introducing logistic growth. Focusing again on the
deterministic situation first, this adds to~\eqref{eq:model} a term
$\alpha \rho (1-\rho/\rho_0)$, where $\alpha$ is the birth rate and
$\rho_0$ is the carrying capacity. Crucially, this breaks detailed
balance, allowing not only steady-state
fluxes~\cite{cates-marenduzzo-pagonabarraga-etal:2010} but also
nonstationary patterns (see below).  When the population dynamics is
much slower than particle motion, $\alpha\ll1$, the bifurcation
diagram shown in Fig.~\ref{fig:projection1} now depicts the projection
of a \textit{slow manifold} $\mathcal{M}$ instead. Diffusive motility
leads to convergence of~$\rho$ to $\mathcal{M}$ on a fast time-scale
$O(1)$ in~$\alpha$, during which the global mean~$\bar \rho$ is almost
conserved. On $\mathcal{M}$ itself, the motion is driven solely by
changing $\bar \rho$ via the slow logistic term on time-scales
$O(\alpha^{-1})$.

Three different regimes can then be identified on varying the carrying
capacity~$\rho_0$. For $\rho_0<\rho_{\text{S}}$ the system remains in
a uniform phase, in which dilute motile bacteria homogeneously cover
the domain at $\rho(x)=\rho_0$~\footnote{Note that for
  $\rho_\text{B}<\rho_0<\rho_\text{S}$, the two phase separated states
  $\rho_{\text{L,R}}$ are destabilized by the $\alpha$ term.}.  At
higher capacities homogeneous solutions cannot exist, and the solution
jumps to one of the stable branches, leading to a dense layer of
immotile bacteria at one of the boundaries, which forms on the fast
time-scale. However, for carrying capacities
$\rho_{\text{S}}<\rho_0<\rho_0^c$, the resulting colony is
destabilized by the birth/death term; two limit cycles then appear.
As soon as a colony emerges at one end of the domain, the bacteria in
it start to slowly die out until the global density drops again to
$\rho_{\text{B}}$. It then disperses rapidly, reverting to a uniform
phase, whose density slowly grows until $\rho_{\text{S}}$ is reached
and the cycle begins anew. Symmetry is broken by weak memory of the
previous cycle, so that the colony always reforms in the same
place. (This also holds with periodic boundary conditions, but the
location is then arbitrary.)  One of these limit cycles is projected
onto the bifurcation diagram in Fig~\ref{fig:projection1} (left), and
its spatio-temporal evolution shown in Fig.~\ref{fig:projection1}
(right, a).

Finally, for $\rho_0>\rho_0^c$, all stable fixed points of the system
comprise a dense colony in coexistence with a planktonic `vapor'. A
steady flux of particles from the vapor balances cell-death within the
colony, so that the macroscopic model is stationary, while the
microscopic dynamics are
not~\cite{cates-marenduzzo-pagonabarraga-etal:2010}.  Such fixed
points correspond to points on stable branches of the slow manifold
$\mathcal{M}$. For $d = 1$ there are two such branches
($\rho_{\text{L,R}}$), with more in $d\ge 2$, corresponding to (say)
colonies located in the corners of a square in $d=2$.  For these
computations and the ones below, we picked $\alpha=10^{-4}$ and
$\delta^2=2\cdot10^{-3}$, which leads to
$\rho_{\text{S}}\approx2.08$, $\rho_{\text{B}}\approx 1.64$, and
$\rho_0^c\approx 6.749$. The complete phase diagram in $\delta$ and
$\rho_0$ is shown in Fig.~\ref{fig:phasediag}.

So far we have neglected the effect of intrinsic fluctuations, both in
the diffusive and the reproductive dynamics. The fluctuations in
diffusion can be formally accounted for by adding in~\eqref{eq:model}
a noise-term $\sqrt{N^{-1}} \nabla \cdot (\sqrt{2\rho D(\rho)} \etab)$,
where $\etab$ is spatio-temporal white noise and $N$ is the typical
number of particles present in the domain $\Omega$.  On the other
hand, the population dynamics can be modeled by a reversible reaction
$A\leftrightarrow A+A$ with forward and backward rates $r_f=\alpha$
and $r_b=\alpha/(N\rho_0)$~\cite{Note1337}.
%\footnote{In a system without other
%  interactions, this would not break detailed balance. But detailed
%  balance under \eqref{eq:model0} would require nonconstant reaction
%  rates.}.
The combined effect is captured by a Markov jump
process. In the limit $N\to\infty$, this gives the logistic growth
term $\alpha\rho(1-\rho/\rho_0)$ considered before, with fluctuations
that are Poisson at each location, scaling again with
$\sqrt{N^{-1}}$~\cite{shwartz-weiss:1995}.
\begin{figure*}[tb]
  \begin{center}
    \hfill
    \includegraphics[width=0.35\linewidth]{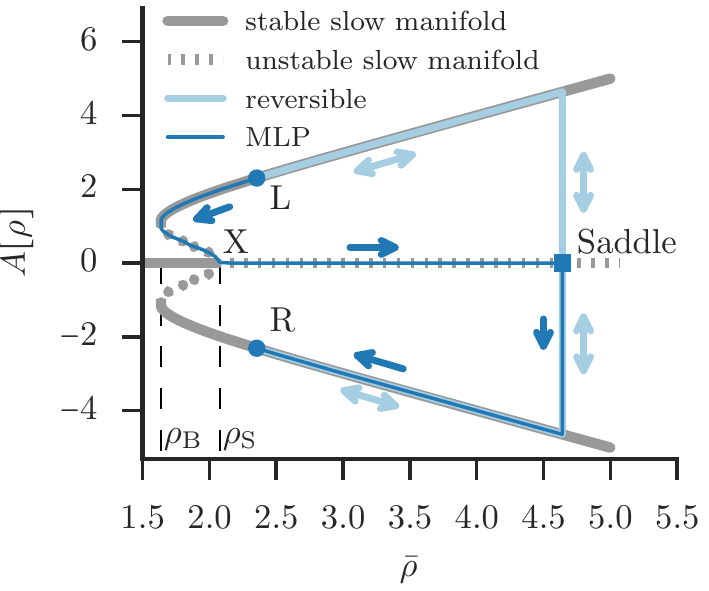}
    \hfill
    \includegraphics[width=0.525\linewidth]{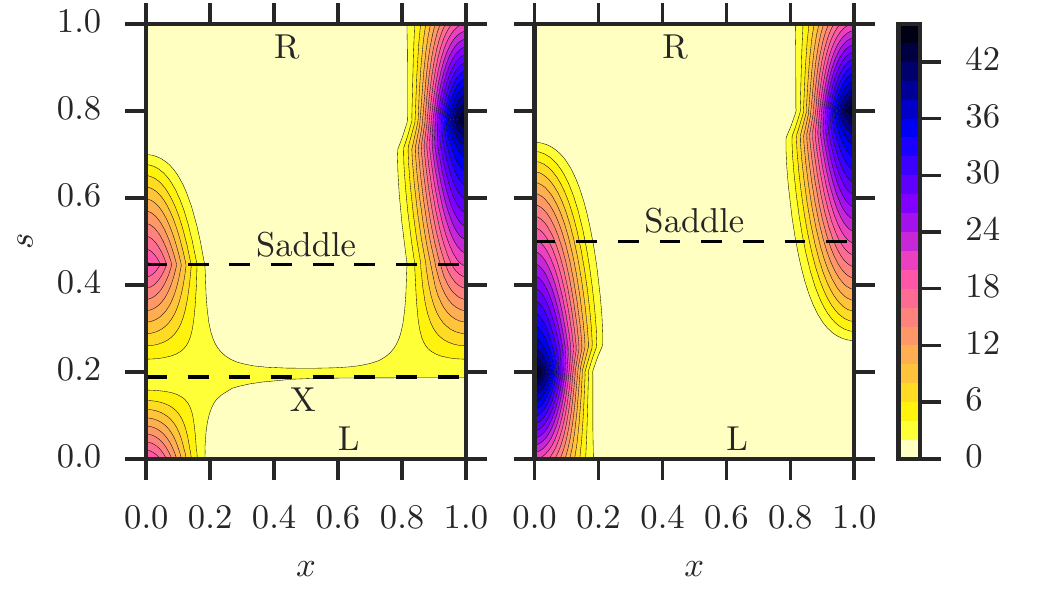}
    \hfill
  \end{center}
  \caption{Transition paths between $\rho_\text{L}$ and
    $\rho_\text{R}$. Here, $x\in[0,1]$ denotes the spatial extend and
    $s\in[0,1]$ the normalized arc-length along each
    trajectory. \emph{Left:} Projection into bifurcation diagram, see
    Fig.~\ref{fig:projection1}. \emph{Center:} MLP from
    $\rho_\text{L}$ to $\rho_\text{R}$. \emph{Right:} Reversible
    transition from $\rho_\text{L}$ to
    $\rho_\text{R}$. \label{fig:transition}}
\end{figure*}

We consider only the weak fluctuation (large $N$) regime which can be
captured by LDT~\cite{freidlin-wentzell:2012}.  This theory predicts
that a noise-driven event occurs, with probability close to $1$, via
the path involving the least unlikely fluctuation able to drive this
event. The resulting most likely path (MLP, also referred to as the
\emph{instanton}) is the minimizer of the action functional
\begin{equation}
  \label{eq:action}
  %I_T(\rho) = \frac12\int_0^T\!dt \int_\Omega\!d\xb\left|\theta(t,\xb;\rho)\right|^2
  I_T[\rho] = \frac12\int_0^T\!dt \int_\Omega\!d\xb \,\theta(\xb,t;\rho) \partial_t \rho(\xb,t)
\end{equation}
which quantifies the likelihood of the `fluctuation'
$\theta(\xb,t;\rho)$.  In our context, $\theta(\xb,t;\rho)$ is related
to $\rho(\xb,t)$~\cite{Note137} via
\begin{equation}
  \label{eq:2}
  \begin{aligned}
    \partial_t \rho &=\nabla \cdot
      (\mathcal{D}_e(\rho) \nabla \rho - \rho D(\rho) \nabla (\delta^2
      \Delta \rho + 2\theta) ) \\
    &\quad + \alpha\rho e^{\theta} - \alpha\rho^2 e^{-\theta}/\rho_0\,.
  \end{aligned}
\end{equation}
The action $I_T$ should be minimized over all paths, and all durations
$T$, consistent with the event of interest, giving $I^* = \inf_{T>0}
\inf_{\rho} I_T[\rho]$. The minimizer is the MLP for the event, whose
rate is then, up to a prefactor, $\exp(-N I^*)$
~\footnote{The numerical details of the minimization are discussed
  in~\cite{heymann-vanden-eijnden:2008,
    grafke-schaefer-vanden-eijnden:2016}.}.

In the regime $\rho_0<\rho_{\text{S}}$, the spatially homogeneous
configuration is stable and LDT predicts that deviations away from it
are exponentially rare and transient. This is no longer true for
$\rho_{\text{S}}<\rho_0<\rho_0^c$: Since part of the limit cycle
tracks $O(\alpha)$ close to the separatrix for times $O(\alpha^{-1})$,
even tiny fluctuations can trigger a crossing of the separatrix into
the other limit cycle. This is consistent with LDT, for which in the
limit $\alpha\to 0$ zero-action minimizers connect these two limit
cycles. Therefore, due to the fluctuations, bacterial colonies
randomly appear either on the left or right wall, disappearing again
after times $O(\alpha^{-1})$. This switching behavior is depicted in
Fig.~\ref{fig:projection1}~(b). With periodic boundary conditions, the
colony instead appears at a random location each cycle, as shown in
Fig.~\ref{fig:projection1}~(c). Note that the time period of the
cycles is not affected significantly by these fluctuations.

For $\rho_0\,>\,\rho_0^c$, on the other hand, $\rho_\text{L}$ and
$\rho_\text{R}$ become meta\-stable; the noise triggers rare and
aperiodic transitions between them.  A projection of the resulting MLP
from $\rho_\text{L}$ to $\rho_\text{R}$ onto the bifurcation diagram
is shown in Fig.~\ref{fig:transition} (left), whereas its actual shape
is shown in Fig.~\ref{fig:transition} (center); that from
$\rho_\text{R}$ to $\rho_\text{L}$ follows by symmetry.  To understand
its features, notice that the slow manifold connects the two stable
fixed points $\rho_\text{L}$ and $\rho_\text{R}$ through the
bifurcation point, and so can be used as a channel to facilitate the
transition. This is indeed correct for $\delta\ll 1$, as the free
energy barrier for a jump between stable branches scales like
$\delta^{-1}$~\cite{grafke-hirschberg-vanden-eijnden:2017}. It is
confirmed by a numerical calculation of the MLP as shown in
Fig.~\ref{fig:transition} (center): The colony of bacteria on the left
first disperses to form a uniform planktonic phase which then attains
the bifurcation point $\rho_\text{X}(\xb)\,=\,\rho_\text{S}$. The system
then follows the unstable branch (separatrix) of $\mathcal{M}$ -- with
two symmetric colonies -- to the saddle point at
$\rho_{\text{Saddle}}(\xb)$, where it enters the basin of attraction
of $\rho_\text{R}$.  Note that for finite $N$, diffusive noise
will push the actual trajectory off the separatrix well before it
reaches the transition state at $\rho_{\text{Saddle}}$. The event rate
is still found by LDT, since the motion after visiting $\rho_\text{X}$
is effectively deterministic and does not contribute to $I^*$.

Let us stress the non-equilibrium nature of this transition: If the
system were in detailed balance, time-reversal symmetry would require
the transition path to follow the deterministic relaxation trajectory
in reverse from $\rho_\text{L}$ to the transition state
$\rho_{\text{Saddle}}$, and then relax deterministically to
$\rho_\text{R}$ (see Fig.~\ref{fig:transition}). Along this
trajectory, the colony initially grows instead of shrinking, bringing
the global average density $\bar \rho$ up to
$\rho_{\text{Saddle}}$. Subsequently, bacteria leave the colony, cross
the low-density region in the center, and accumulate on the opposite
wall, keeping $\bar\rho$ constant. This part of the transition happens
diffusively causing the system to leave the slow manifold. After
passing through $\rho_{\text{Saddle}}$, the reversible transition
necessarily coincides with the true MLP.

In summary, we have analyzed a minimal model for the fluctuating
dynamics of self-organization among motile bacteria in a finite
domain, in the presence of birth and death processes that are slow
compared to diffusive time-scales.  For
$\rho_{\text{S}}<\rho_0<\rho_0^c$, a bacterial colony is present;
fluctuations allow it to explore the domain by random relocation at
regular intervals.  This exploration remains possible at higher
carrying capacities, $\rho_0>\rho_0^c$, where fluctuations now induce
exponentially rare, aperiodic relocations. Such transitions are
inherently out-of-equilibrium: their most likely path differs
significantly from the one followed in systems with time-reversible
dynamics. In our model, relocation of a colony is triggered by a slow
decline in local population density, followed by rapid dispersal and
re-formation elsewhere, rather than by progressive migration between
old and new sites.  The intermediate state has no colonies in the
first case and two in the second -- alternatives that should be
clearly distinguishable experimentally. Similar considerations may
apply to other organisms that alternate between nomadic and
cooperative lifestyles. The formalism readily applies to a
higher-dimensional setup, e.g.~bacterial colonies in the corners of a
two-dimensional rectangular domain, and a slow manifold with four
stable branches.

\textbf{Acknowledgments.} We thank T.~Sch\"afer for his help with the
numerical scheme and A.~Donev, O.~Hirschberg, and C.~Nardini for
interesting discussions. MEC is funded by the Royal Society.  EVE is
supported in part by the Materials Research Science and Engineering
Center (MRSEC) program of the National Science Foundation (NSF) under
award number DMR-1420073 and by NSF under award number DMS-1522767.

\nocite{bertini-de_sole-gabrielli-etal:2015}

%\bibliographystyle{apsrev4-1}
%\bibliography{bib}

\clearpage


\section*{Supplementary material}

# SUPPLEMENTAL MATERIAL: SPATIOTEMPORAL SELF-ORGANIZATION OF FLUCTUATING BACTERIAL COLONIES

TOBIAS GRAFKE, MICHAEL E. CATES, AND ERIC VANDEN-EIJNDEN

Below we give some additional details about the derivation of the equations given in main text, and in particular explain: (i) how to obtain a spatially-continuous model from a microscopic description of the microorganisms based on a lattice gas model, and (ii) how to account for the effects of fluctuations at continuous level using a large deviation principle (LDP).

1. **Motile microorganisms modeled as lattice gas.** We model the behavior of spatially extended bacterial colonies as a Markov jump process (MJP) of particles on a one-dimensional lattice with $L$ sites of physical width $h = 1/L$, i.e. such that the extent of the physical domain $\Omega$ is normalized to $\Omega = [0,1]$ – the generalization to multidimensional domains is straightforward but makes the notations more cumbersome so we avoid it here. We assume that there exist two sub-populations moving left or right, and denote by $n_i^-$ the number of left-moving particles at site $i$ and by $n_i^+$ the number of right-moving particles at site $i$. We use the vectors $\boldsymbol{n}^\pm = (n_1^\pm, \ldots, n_L^\pm) \in \mathbb{N}^L$ to describe the complete configuration of all the particles in the domain. We assume that the rates of hopping from site $i$ to site $i-1$ (for the left-moving particles) or site $i$ to site $i+1$ (for the right-moving particles) are the same and depend on the total number of particles at site $i$: $n_i = n_i^- + n_i^+$ (*quorum sensing*). Denoting this common rate by $v(n_i)$, the statistical evolution of any observable $f(t, \boldsymbol{n}_0^-, \boldsymbol{n}_0^+) \equiv \mathbb{E}\left(\phi(\boldsymbol{n}^-(t), \boldsymbol{n}^+(t)) | \boldsymbol{n}^-(0) = \boldsymbol{n}_0^-, \boldsymbol{n}^+(0) = \boldsymbol{n}_0^+\right)$ can be described by the backward Kolmogorov equation (BKE),

$$\partial_t f(t, \boldsymbol{n}^-, \boldsymbol{n}^+) = L_{\text{prop}} f(t, \boldsymbol{n}^-, \boldsymbol{n}^+), \qquad f(0, \boldsymbol{n}^-, \boldsymbol{n}^+) = \phi(\boldsymbol{n}^-, \boldsymbol{n}^+), \tag{SM.1}$$

where $L_{\text{prop}}$ is the generator of the MJP (we use the subscript 'prop' to distinguish this propulsion part of the generator with the reproduction part considered below):

$$\begin{aligned}
L_{\text{prop}} f(t, \boldsymbol{n}^-, \boldsymbol{n}^+) &= \sum_{i=1}^{L} n_i^+ v(n_i) \left(f(t, \boldsymbol{n}^-, \boldsymbol{n}^+ + \boldsymbol{e}_i^+) - f(t, \boldsymbol{n}^-, \boldsymbol{n}^+)\right) &&\text{(right jumps)} \\
&+ \sum_{i=1}^{L} n_i^- v(n_i) \left(f(t, \boldsymbol{n}^- + \boldsymbol{e}_i^-, \boldsymbol{n}^+) - f(t, \boldsymbol{n}^-, \boldsymbol{n}^+)\right) &&\text{(left jumps)} \\
&+ \tau^{-1} \sum_{i=1}^{L} n_i^- \left(f(t, \boldsymbol{n}^- - \boldsymbol{e}_i, \boldsymbol{n}^+ + \boldsymbol{e}_i) - f(t, \boldsymbol{n}^-, \boldsymbol{n}^+)\right) &&\text{(tumbles right to left)} \\
&+ \tau^{-1} \sum_{i=1}^{L} n_i^+ \left(f(t, \boldsymbol{n}^- + \boldsymbol{e}_i, \boldsymbol{n}^+ - \boldsymbol{e}_i) - f(t, \boldsymbol{n}^-, \boldsymbol{n}^+)\right) &&\text{(tumbles left to right)}
\end{aligned} \tag{SM.2}$$

Here $\tau$ defines the tumbling rate and

$$\boldsymbol{e}_i^+ = (0, \ldots, 0, \underbrace{-1}_{i}, \underbrace{1}_{i+1}, 0, \ldots, 0), \quad \boldsymbol{e}_i^- = (0, \ldots, 0, \underbrace{1}_{i-1}, \underbrace{-1}_{i}, 0, \ldots, 0), \quad \boldsymbol{e}_i = (0, \ldots, 0, \underbrace{1}_{i}, 0, \ldots, 0). \tag{SM.3}$$

The particular form of the rate $v$ depends on physical considerations on the model of the motility of the random walkers. Later on, we will add non-locality to the rate to realize quorum sensing. In the next section, we are going to make this model continuous, first by considering the large number of particle limit and then by taking the continuous limit in space.

1.1. *Derivation of the continuous model from the microscopic model.* Let $N$ be a typical number of particles in the domain. Denote by $\rho_i^\pm = L n_i^\pm / N$ the particle density at site $i$. This choice would yield a normalization $h \sum_i \rho_i = 1$ if we had $\sum_i n_i = N$, but since we later add reproduction to the model the total number of particles is not conserved and $N$ is

---

merely a scaling parameter. By writing $\epsilon = L/N$, and redefining the functions $f$ and $v$ appropriately, the BKE becomes

$$\partial_t f(t, \boldsymbol{\rho}, \boldsymbol{\eta}) = \frac{1}{\epsilon} \sum_{i=0}^{L} \rho_i^+ v(\rho_i) \left( f(t, \boldsymbol{\rho}^-, \boldsymbol{\rho}^+ + \epsilon \boldsymbol{e}_i^+) - f(t, \boldsymbol{\rho}^-, \boldsymbol{\rho}^+) \right) \quad \text{(right jumps)}$$
$$+ \frac{1}{\epsilon} \sum_{i=0}^{L} \rho_i^- v(\rho_i) \left( f(t, \boldsymbol{\rho}^- + \epsilon \boldsymbol{e}_i^-, \boldsymbol{\rho}^+) - f(t, \boldsymbol{\rho}^-, \boldsymbol{\rho}^+) \right) \quad \text{(left jumps)}$$
$$+ \frac{1}{\epsilon \tau} \sum_{i=1}^{L} \rho_i^- \left( f(t, \boldsymbol{\rho}^- - \epsilon \boldsymbol{e}_i, \boldsymbol{\rho}^+ + \epsilon \boldsymbol{e}_i) - f(t, \boldsymbol{\rho}^-, \boldsymbol{\rho}^+) \right) \quad \text{(tumbles right to left)}$$
$$+ \frac{1}{\epsilon \tau} \sum_{i=1}^{L} \rho_i^+ \left( f(t, \boldsymbol{\rho}^- + \epsilon \boldsymbol{e}_i, \boldsymbol{\rho}^+ - \epsilon \boldsymbol{e}_i) - f(t, \boldsymbol{\rho}^-, \boldsymbol{\rho}^+) \right). \quad \text{(tumbles left to right)}$$
(SM.4)

In order to consider the large particle limit, we let $\epsilon \to 0$, that is, we assume that the number of particles per cell is large. At the same time, we make a WKB ansatz, i.e. we set $f(t, \boldsymbol{\rho}^-, \boldsymbol{\rho}^+) = \exp(S(t, \boldsymbol{\rho}^-, \boldsymbol{\rho}^+)/\epsilon)$. This will allow us to derive a large deviation principle (LDP) for the stochastic system [21, 30], while avoiding introducing stochastic partial differential equations (SPDEs), which from a mathematical perspective are ill-defined in the continuous space limit $h \to 0$ due to the multiplicative spatio-temporal white noise and the Poisson-process for the reaction terms. To leading order in $\epsilon$ the function $S$ satisfies the following Hamilton-Jacobi equation (HJE):

$$\partial_t S(t, \boldsymbol{\rho}^-, \boldsymbol{\rho}^+) = H(\boldsymbol{\rho}^-, \boldsymbol{\rho}^+, \nabla_{\boldsymbol{\rho}^-} S, \nabla_{\boldsymbol{\rho}^+} S) \tag{SM.5}$$

with the Hamiltonian

$$H(\boldsymbol{\rho}^-, \boldsymbol{\rho}^+, \boldsymbol{\theta}^-, \boldsymbol{\theta}^+) = \sum_{i=1}^{L} \left( \rho_i^- v(\rho_i) \left( e^{\theta_{i-1}^- - \theta_i^-} - 1 \right) + \rho_i^+ v(\rho_i) \left( e^{\theta_{i+1}^+ - \theta_i^+} - 1 \right) \right) \quad \text{(jumps)}$$
$$+ \tau^{-1} \sum_{i=0}^{L} \left( \rho_i^- \left( e^{\theta_i^+ - \theta_i^-} - 1 \right) + \rho_i^+ \left( e^{-\theta_i^+ + \theta_i^-} - 1 \right) \right). \quad \text{(tumbles)}$$
(SM.6)

Next, we take the continuous space limit, i.e. let $L \to \infty$ or equivalently $h \to 0$. To this end, we define $x_i = ih$ and the spatially dependent functions $\rho^\pm(x)$ through $\rho^\pm(x_i) = \rho_i^\pm$ and similarly $\theta^\pm(x)$ through $\theta^\pm(x_i) = \theta_i^\pm$. If we rescale time according to $t \to ht$ and rescale $u \to v/h$, in the limit as $h \to 0$ the sums converges to integrals, and the HJE above converges to the functional HJE

$$\partial_t \mathcal{S}(t, \rho^-, \rho^+) = \mathcal{H}(\rho^-, \rho^+, \delta \mathcal{S}/\delta \rho^-, \delta \mathcal{S}/\delta \rho^+) \tag{SM.7}$$

with the continuous Hamiltonian

$$\mathcal{H}[\rho^-, \rho^+, \theta^-, \theta^+] = \int_\Omega \left( -\rho^- v(\rho) \partial_x \theta^- + \rho^+ v(\rho) \partial_x \theta^+ \right) dx \quad \text{(jumps)}$$
$$+ \tau^{-1} \int_\Omega \left( \rho^- (e^{-\theta^- + \theta^+} - 1) + \rho^+ (e^{\theta^- - \theta^+} - 1) \right) dx. \quad \text{(tumbles)}$$
(SM.8)

The functional HJE (SM.7) can be given a precise meaning via its associated Hamilton's equations (which are partial differential equation (PDEs) for $\rho^\pm$ and $\theta^\pm$), as described below. The HJE also permits to derive a law of large number (LLN) and a large deviation principle (LDP), which in turns allow one to describe the effect of fluctuations in the system in the large particle number and spatially-continuous limits, see Sec. 4. Here we will derive the corresponding equations in a regime where the tumbling occurs on a fast time scale, i.e. when $\tau \to 0$.

1.2. *Averaging out the tumbling.* In the fast tumbling limit when $\tau \to 0$, we can obtain a closed effective HJE for the total density of particles, since differences between the densities of left and right-movers are averaged out after times order $O(\tau)$. To derive this effective equation, we rescale time as $t \to t/\tau$ and introduce the variables $\rho = \rho^+ + \rho^-$ and $\eta = \eta^+ - \eta^-$, in terms of which the functional HJE (SM.7) becomes

$$\partial_t \mathcal{S}(t, \rho, \eta) = \mathcal{H}(\rho, \eta, \delta \mathcal{S}/\delta \rho, \delta \mathcal{S}/\delta \eta) \tag{SM.9}$$

for the Hamiltonian

$$\mathcal{H}[\rho, \eta, \theta, \zeta] = \int_\Omega \left( \eta v(\rho) \partial_x \theta + \rho v(\rho) \partial_x \zeta \right) dx$$
$$+ \tau^{-1} \int_\Omega \left( \tfrac{1}{2} \rho \left( e^{2\zeta} + e^{-2\zeta} - 2 \right) + \tfrac{1}{2} \eta \left( e^{-2\zeta} - e^{2\zeta} \right) \right) dx.$$
(SM.10)

This is equivalent to making the following canonical transformation on the Hamilton's equations associated with (SM.7)

$$\rho = \rho^+ + \rho^-, \quad \eta = \rho^+ - \rho^-, \quad \theta = \tfrac{1}{2}(\theta^+ + \theta^-), \quad \zeta = \tfrac{1}{2}(\theta^+ - \theta^-), \tag{SM.11}$$

which transform them into (these are Hamilton's equations associated with (SM.9))

$$\begin{cases} \partial_t \rho = \delta H/\delta\theta = -\tau^{-1}\partial_x(v(\rho)\eta) \\ \partial_t \eta = \delta H/\delta\zeta = -\tau^{-1}\partial_x(v(\rho)\rho) + \tau^{-2}\rho\left(e^{2\zeta} - e^{-2\zeta}\right) - \tau^{-2}\eta\left(e^{2\zeta} + e^{-2\zeta}\right) \\ \partial_t \theta = -\delta H/\delta\rho = -\tau^{-1}(\rho v'(\rho) + v(\rho))\partial_x\zeta - \tau^{-1}\eta v'(\rho)\partial_x\theta - \frac{1}{2}\tau^{-2}\left(e^{2\zeta} + e^{-2\zeta} - 2\right) \\ \partial_t \zeta = -\delta H/\delta\eta = -\tau^{-1}v(\rho)\partial_x\theta - \frac{1}{2}\tau^{-2}\left(e^{-2\zeta} - e^{2\zeta}\right). \end{cases} \qquad \text{(SM.12)}$$

We look for solutions of these equations scaling like $\eta \sim O(\tau), \zeta \sim O(\tau)$, which is the correct scaling to get a limiting set of equations when tumbling is fast and the difference of populations of left and right-movers scales like $\tau$. To this end, we replace $\eta \to \tau\eta$ and $\zeta \to \tau\zeta$ in (SM.12), expand in powers of $\tau$, and collect terms of the same order. To leading order, $O(\tau^{-1})$, the second and fourth equation in (SM.12) give

$$\begin{cases} 0 = -\partial_x(v(\rho)\rho) + 4\rho\zeta - 2\eta \\ 0 = -v(\rho)\partial_x\theta + 2\zeta \end{cases} \qquad \text{(SM.13)}$$

which implies that

$$\begin{cases} \eta = -\frac{1}{2}\partial_x(v(\rho)\rho) + 2\rho\zeta = -\frac{1}{2}\partial_x(v(\rho)\rho) + \rho v(\rho)\partial_x\theta \\ \zeta = \frac{1}{2}v(\rho)\partial_x\theta. \end{cases} \qquad \text{(SM.14)}$$

By re-inserting (SM.14) into (SM.9) and using (SM.10), we obtain the following limiting HJE on the time scale where the tumbling has been averaged out

$$\partial_t \mathcal{S}(t, \rho) = \mathcal{H}(\rho, \delta\mathcal{S}/\delta\rho) \qquad \text{(SM.15)}$$

with the limiting Hamiltonian

$$\mathcal{H}[\rho, \theta] = \frac{1}{2}\int_\Omega \left(\partial_x(v(\rho)\partial_x(v(\rho)\rho))\theta + \rho v^2(\rho)(\partial_x\theta)^2\right) dx \qquad \text{(SM.16)}$$

If we define $D(\rho) = \frac{1}{2}v^2(\rho)$ so that $v(\rho)v'(\rho) = D'(\rho)$, this limiting Hamiltonian can be rewritten as

$$\mathcal{H}[\rho, \theta] = \int_\Omega \left(\partial_x((D(\rho) + \frac{1}{2}\rho D'(\rho))\partial_x\rho)\theta + \rho D(\rho)(\partial_x\theta)^2\right) dx, \qquad \text{(SM.17)}$$

which yields an effective diffusivity

$$\mathcal{D}_e(\rho) = D(\rho) + \frac{1}{2}\rho D'(\rho). \qquad \text{(SM.18)}$$

The corresponding Hamilton's equations are also the limiting equations obtained by inserting (SM.14) into the first and third equations in (SM.12). They read

$$\begin{cases} \partial_t \rho = \partial_x\left(\mathcal{D}_e(\rho)\partial_x\rho\right) - 2\partial_x\left(\rho D(\rho)\partial_x\theta\right) \\ \partial_t \theta = \mathcal{D}_e(\rho)\partial_x^2\theta + (D(\rho) + \rho D'(\rho))(\partial_x\theta)^2. \end{cases} \qquad \text{(SM.19)}$$

Note that these equations are ill-posed if $\mathcal{D}_e(\rho) < 0$. This is precisely the situation of interest to us, since it is the one that leads to MIPS. In order to regularize (SM.15) and the associated (SM.19), we introduce next a physical length scale over which the particles feel each other.

2. **Quorum sensing and non-locality.** Up to now, the propulsion speed of the active particles, $v(\rho)$, was set to depend on the local particle density $\rho$. In this section we additionally want to modify the quorum sensing by introducing a physical length scale $\delta$ over which the active walkers feel each other's influence. In effect, the jump rates $v$ are taken to depend not only of the number of particles at the local lattice site, but also of that number at neighboring sites. To achieve this on the level of the lattice gas, we choose a fraction $\delta$ of the the physical domain $[0, 1]$ and introduce a set of coefficients $\{a_k^\delta\}_{k\in\mathbb{N}}$ with $a_k^\delta = a_{-k}^\delta$ that decay to zero fast for $|k| > \delta L$. We then make the jump rate of particles at the site $i$ to be $v(n_i^\delta)$ instead of $v(n_i)$, where

$$n_i^\delta = \sum_{j=0}^L h\, a_{i-j}^\delta\, n_j. \qquad \text{(SM.20)}$$

To take the continuous limit $h \to 0$, we define a mollifier $\phi^\delta(x)$ through $\phi^\delta(x_i) = a_i^\delta$ and denote $\rho_i^\delta = Ln_i^\delta/N$ to rewrite the equation above as

$$\rho_i^\delta = \sum_{j=1}^L ha_{i-j}^\delta \rho_j = \sum_{j=1}^L h\phi^\delta(x_i - x_j)\rho(x_j) \qquad \text{(SM.21)}$$

In the limit $h \to 0$ this Riemann sum converges to

$$\rho^\delta(x) \equiv \int_\Omega \phi^\delta(x - y)\rho(y)\, dy = \int_\Omega \phi^\delta(y - x)\rho(y)\, dy, \qquad \text{(SM.22)}$$

3

i.e. a convolution of the density field $\rho(x)$ with a mollifier of scale $\delta$, and the jump rates are now taken as $v(\rho^\delta)$. With this modification, the Hamiltonian (SM.17) to be used in the HJE (SM.15) then becomes

$$\mathcal{H}[\rho, \theta] = \int_\Omega \left(\partial_x(D(\rho^\delta)\partial_x \rho + \tfrac{1}{2}\rho D'(\rho^\delta)\partial_x \rho^\delta)\,\theta + \rho D(\rho^\delta)(\partial_x \theta)^2\right) dx. \tag{SM.23}$$

Unlike (SM.19) the Hamilton's equations associated with the HJE with this Hamiltonian are well-posed even if $\mathcal{D}_e(\rho) < 0$.

2.1. *Thermodynamic mapping of the diffusive dynamics.* Let us investigate the conditions necessary to map the stochastic process defined above to a thermal system that obeys *detailed balance* (aka microscopic reversibility). Detailed balance is fulfilled if all forces derive from a free energy $E[\rho]$, and diffusion and mobility operators obey Einstein's relations, i.e. we can write the Hamiltonian in the form

$$\mathcal{H}[\rho, \theta] = -\langle \mathcal{M}[\rho]\delta E[\rho]/\delta\rho, \theta \rangle + \langle \theta, \mathcal{M}[\rho]\theta \rangle, \tag{SM.24}$$

for a free energy $E[\rho]$ and a mobility operator $\mathcal{M}[\rho]$, where $\langle \cdot, \cdot \rangle$ is the $L^2(\Omega)$ inner product [39]. It is easy to see that (SM.23) can be written in the form of (SM.24) if (i) we set

$$M[\rho]\xi = -\partial_x(\rho D(\rho^\delta)\partial_x \xi) \tag{SM.25}$$

for the mobility and (ii) we choose

$$E[\rho] = \int_\Omega (\rho \log \rho - \rho)\, dx + F_{\text{ex}}[\rho] \tag{SM.26}$$

for the energy, with the excess free energy $F_{\text{ex}}$ defined through

$$\frac{\delta F_{\text{ex}}[\rho]}{\delta \rho} = \tfrac{1}{2} \log(D(\rho^\delta)). \tag{SM.27}$$

This procedure is referred to as *thermodynamic mapping* in [9]. If it can be achieved, then the stationary solution to the HJE, that is, the solution to

$$0 = H[\rho, \delta\mathcal{S}/\delta\rho] \tag{SM.28}$$

is simply $\mathcal{S}[\rho] = E[\rho]$. This in turns is consistent with the equilibrium measure of the process being proportional to $\exp(-\epsilon^{-1}E[\rho])$ (i.e. to te Boltzmann-Gibbs measure), as expected for a process in detailed-balance, but it gives a precise meaning within large deviation theory (LDT) to this measure which may be ill-defined in the continuous setting.

Remembering that $\rho^\delta(x)$ is given by (SM.22), it turns out that only a specific choice of $D(\rho^\delta)$ can fulfill the equation (SM.27), i.e. allow $\log(D(\rho^\delta))$ to be written as a functional gradient. This can be demonstrated by realizing that the functional second derivative has to be symmetric, i.e. we must have

$$\frac{\delta}{\delta\rho(x)}\left(\frac{\delta F_{\text{ex}}}{\delta\rho(y)}\right) = \frac{\delta}{\delta\rho(y)}\left(\frac{\delta F_{\text{ex}}}{\delta\rho(x)}\right).$$

Since, by (SM.27),

$$\frac{\delta}{\delta\rho(x)}\left(\frac{\delta F_{\text{ex}}}{\delta\rho(y)}\right) = \tfrac{1}{2}\frac{\delta}{\delta\rho(x)}\log(D(\rho^\delta(y))) = \tfrac{1}{2}\frac{D'(\rho^\delta(y))}{D(\rho^\delta(y))}\frac{\delta\rho^\delta(y)}{\delta\rho(x)} = \tfrac{1}{2}\frac{D'(\rho^\delta(y))}{D(\rho^\delta(y))}\phi^\delta(x - y)$$

we see that we must have $D'(\rho^\delta)/D(\rho^\delta)$ = cst., or equivalently,

$$D(\rho^\delta) = e^{-\lambda \rho^\delta}, \tag{SM.29}$$

i.e. exponential decay of the density dependent diffusivity. We choose $\lambda = 1$. For this choice, we have

$$F_{\text{ex}}[\rho] = -\tfrac{1}{4}\lambda \int_\Omega \rho \rho^\delta\, dx \quad \text{and} \quad E[\rho] = \int_\Omega \left(\rho \log \rho - \rho - \tfrac{1}{4}\rho\rho^\delta\right) dx. \tag{SM.30}$$

In the following, for most concrete computations, we will therefore assume the form (SM.29). Note that [18] does assume the exponential form (SM.29) but combines this with a phenomenological nonlocal/gradient contribution that differs from the one found here by explicit coarse-graining. Our approach is clearly preferable.

2.2. *Regularizing surface term.* Due to the way density-dependent motility is modeled in the microscopic models, all effective equations obtained above are non-local integro-differential equations. Under the assumption that the length scale associated with the quorum sensing is small compared to the length of the domain, $\delta \ll 1$, it is possible to expand in $\delta$ to arrive at purely local models, where the nonlocal mechanism is transformed into a fourth-order regularizing surface tension term. In contrast to [18], the regularizing term introduced in this section is consistent with the thermodynamic mapping discussed above.

With the choice $\phi^\delta(x) = \frac{1}{2\sqrt{2\pi}\delta} e^{-\frac{1}{8}x^2/\delta^2}$ or in Fourier space $\hat{\phi}^\delta(k) = e^{-2k^2\delta^2}$, we can expand the convolution for $\delta \ll 1$ to obtain an expansion

$$\hat{\phi}^\delta(k) = 1 - 2k^2\delta^2 + \mathcal{O}(\delta^4).$$

As a consequence,

$$\rho^\delta(x) = \rho(x) + 2\delta^2 \partial_x^2 \rho(x) + \mathcal{O}(\delta^4),$$

which implies that, to $\mathcal{O}(\delta^4)$,

$$E[\rho] = \int_\Omega \left( \rho \log \rho - \rho - \tfrac{1}{4}\rho^2 + \tfrac{1}{2}\delta^2 |\partial_x \rho|^2 \right) dx \tag{SM.31}$$

as the effective free energy. Similarly, we have $D(\rho^\delta) = D(\rho) + 2\delta^2 D'(\rho)\partial_x^2 \rho + \mathcal{O}(\delta^4)$, which implies that, to leading order in $\delta$, the mobility is given by

$$M[\rho]\xi = \partial_x(\rho D(\rho)\partial_x \xi). \tag{SM.32}$$

Using (SM.31) and (SM.32) in (SM.24) leads to the Hamiltonian:

$$\mathcal{H}[\rho, \theta] = \int_\Omega \theta \partial_x \left( \mathcal{D}_e(\rho)\partial_x \rho - \rho D(\rho)(\delta^2 \partial_x^2 \rho + \theta) \right) dx. \tag{SM.33}$$

The corresponding LLN obtained by setting $\theta = 0$ in $\partial_t \rho = \delta \mathcal{H}/\delta \theta$ recovers equation (1) in the main text.

3. **Birth and death.** Additionally, bacteria reproduce and compete for resources. This mechanism can be represented locally as a continuous time MJP with two reactions: with a rate $\alpha$, a particle, regardless of its orientation, splits into two, modeling the reproduction of the bacteria. With a rate $\alpha n_i/n_0$, one particle is annihilated, modeling the competition of bacteria for resources. Note that left and right-movers compete for the same resource with total carrying capacity $n_0$. Here, $\alpha$ defines the time scale of the reproduction. Going back to the microscopic description in terms of $\boldsymbol{n}^\pm$, the reproduction part of the MJP is characterized by adding the following component to the generator (SM.2) entering the BKE (SM.1)

$$\begin{aligned}
L_{\text{rep}} f(t, \boldsymbol{n}^-, \boldsymbol{n}^+) &= \alpha \sum_{i=0}^{L} \left( n_i^+ \left( f(t, \boldsymbol{n}^-, \boldsymbol{n}^+ + \boldsymbol{e}_i) - f(t, \boldsymbol{n}^-, \boldsymbol{n}^+) \right) \right) && \text{(birth of right-mover)} \\
&+ \alpha \sum_{i=0}^{L} \left( n_i^- \left( f(t, \boldsymbol{n}^- + \boldsymbol{e}_i, \boldsymbol{n}^+) - f(t, \boldsymbol{n}^-, \boldsymbol{n}^+) \right) \right) && \text{(birth of left-mover)} \\
&+ \alpha \sum_{i=0}^{L} \left( \frac{n_i^+(n_i - 1)}{n_0} \left( f(t, \boldsymbol{n}^-, \boldsymbol{n}^+ - \boldsymbol{e}_i) - f(t, \boldsymbol{n}^-, \boldsymbol{n}^+) \right) \right) && \text{(death of right-mover)} \\
&+ \alpha \sum_{i=0}^{L} \left( \frac{n_i^-(n_i - 1)}{n_0} \left( f(t, \boldsymbol{n}^- - \boldsymbol{e}_i, \boldsymbol{n}^+) - f(t, \boldsymbol{n}^-, \boldsymbol{n}^+) \right) \right) && \text{(death of left-mover)}
\end{aligned} \tag{SM.34}$$

where the factors $(n_i - 1)$ in the death terms appear because a particle does not contriubute to its own competition and at least two particles need to meet in order for annihilation to happen.

The complete process is therefore described by the BKE

$$\partial_t f(t, \boldsymbol{n}^-, \boldsymbol{n}^+) = L_{\text{rep}} f(t, \boldsymbol{n}^-, \boldsymbol{n}^+) + L_{\text{prop}} f(t, \boldsymbol{n}^-, \boldsymbol{n}^+), \tag{SM.35}$$

that takes into account the stochastic behavior of particles on a lattice subject to reproduction, competition, and active propulsion.

Proceeding as before, we let $\rho_i^\pm = Ln_i^\pm/N$ and similarly $\rho_0 = Ln_0/N$ to obtain

$$L_{\text{rep}} f(t, \boldsymbol{\rho}^-, \boldsymbol{\rho}^+) = \frac{\alpha}{\epsilon} \sum_{i=0}^{L} \left( \rho_i^+ \left( f(t, \boldsymbol{\rho}^-, \boldsymbol{\rho}^+ + \epsilon \boldsymbol{e}_i) - f(t, \boldsymbol{\rho}^-, \boldsymbol{\rho}^+) \right) \right) \quad \text{(birth of right-mover)}$$
$$+ \frac{\alpha}{\epsilon} \sum_{i=0}^{L} \left( \rho_i^- \left( f(t, \boldsymbol{\rho}^- + \epsilon \boldsymbol{e}_i, \boldsymbol{\rho}^+) - f(t, \boldsymbol{\rho}^-, \boldsymbol{\rho}^+) \right) \right) \quad \text{(birth of left-mover)}$$
$$+ \frac{\alpha}{\epsilon} \sum_{i=0}^{L} \left( \frac{\rho_i^+(\rho_i - 1)}{\rho_0} \left( f(t, \boldsymbol{\rho}^-, \boldsymbol{\rho}^+ - \epsilon \boldsymbol{e}_i) - f(t, \boldsymbol{\rho}^-, \boldsymbol{\rho}^+) \right) \right) \quad \text{(death of right-mover)}$$
$$+ \frac{\alpha}{\epsilon} \sum_{i=0}^{L} \left( \frac{\rho_i^-(\rho_i - 1)}{\rho_0} \left( f(t, \boldsymbol{\rho}^- - \epsilon \boldsymbol{e}_i, \boldsymbol{\rho}^+) - f(t, \boldsymbol{\rho}^-, \boldsymbol{\rho}^+) \right) \right) \quad \text{(death of left-mover)}$$

(SM.36)

which in the limit $\epsilon \to 0$, using the WKB approximation, yields the Hamiltonian

$$H_{\text{rep}}(\boldsymbol{\rho}^-, \boldsymbol{\rho}^+, \boldsymbol{\theta}^-, \boldsymbol{\theta}^+) = \alpha \sum_{i=0}^{N} \left( \rho_i^+ \left( e^{\theta_i^+} - 1 \right) + \rho_i^- \left( e^{\theta_i^-} - 1 \right) \right) \quad \text{(births)}$$
$$+ \alpha \sum_{i=0}^{N} \left( \rho_i^+ \rho_i \left( e^{-\theta_i^+} - 1 \right)/\rho_0 + \rho_i^- \rho_i \left( e^{-\theta_i^-} - 1 \right)/\rho_0 \right) \quad \text{(deaths)}$$

(SM.37)

so that in the spatial continuum limit, $h \to 0$,

$$\mathcal{H}_{\text{rep}}[\rho^-, \rho^+, \theta^-, \theta^+] = \alpha \int_\Omega \left( \rho^+ \left( e^{\theta^+} - 1 \right) + \rho^- \left( e^{\theta^-} - 1 \right) \right) dx \quad \text{(births)}$$
$$+ \alpha \int_\Omega \left( \rho\rho^+ \left( e^{-\theta^+} - 1 \right)/\rho_0 + \rho\rho^- \left( e^{-\theta^-} - 1 \right)/\rho_0 \right) dx. \quad \text{(deaths)}$$

(SM.38)

Taking again the canonical transformation (SM.11) gives

$$\mathcal{H}_{\text{rep}}[\rho, \eta, \theta, \zeta] = \int_\Omega \left( \tfrac{1}{2}(\rho + \eta)\left(e^{\theta+\zeta} - 1\right) + \tfrac{1}{2}(\rho - \eta)\left(e^{\theta-\zeta} - 1\right) \right) dx$$
$$+ \int_\Omega \left( \tfrac{1}{2}\rho(\rho + \eta)\left(e^{-(\theta+\zeta)} - 1\right)/\rho_0 + \tfrac{1}{2}\rho(\rho - \eta)\left(e^{-(\theta-\zeta)} - 1\right)/\rho_0 \right) dx.$$

(SM.39)

To leading order with the above scaling assumptions, and rescaling $\alpha \to \tau\alpha$, the full Hamiltonian $\mathcal{H} = \mathcal{H}_{\text{prop}} + \mathcal{H}_{\text{rep}}$ finally reads

$$\mathcal{H}[\rho, \theta] = \int_\Omega \left( \theta \partial_x \left( \mathcal{D}_e \partial_x \rho - \rho D(\rho) \partial_x (\delta^2 \partial_x^2 \rho + \theta) \right) + \alpha\rho(e^\theta - 1) + \alpha \frac{\rho^2}{\rho_0}(e^{-\theta} - 1) \right) dx. \quad \text{(SM.40)}$$

4. **Large deviation principle.** The rate function $I_T[\rho]$ for a LDP is usually written in terms of the *Lagrangian* $\mathcal{L}[\rho, \partial_t \rho]$ [21] via

$$I_T[\rho] = \int_0^T \mathcal{L}[\rho, \partial_t \rho] \, dt. \quad \text{(SM.41)}$$

Since the Fenchel-Legendre transform of the full Hamiltonian $\mathcal{H}$ of (SM.40) to obtain $\mathcal{L}$ is very unwieldy, instead we choose to write

$$I_T[\rho] = \int_0^T \mathcal{L}[\rho, \partial_t \rho] \, dt = \sup_\theta \int_0^T \left( \int_\Omega \theta \partial_t \rho \, dx - \mathcal{H}[\rho, \theta] \right) dt = \sup_\theta \int_0^T \int_\Omega \theta \partial_t \rho \, dx \, dt \quad \text{(SM.42)}$$

in the main text's equation (4), where in the last step we used the fact that $\mathcal{H}[\rho, \theta] \to 0$ in the limit $T \to \infty$. This amounts to solving the Hamilton's equations of the associated stochastic field theory, instead of minimizing the action functional explicitly. The *instanton* equation, corresponding to equation (5) in the main text,

$$\partial_t \rho = \frac{\delta \mathcal{H}[\rho, \theta]}{\delta \theta} = \partial_x (\mathcal{D}_e \partial_x \rho - \rho D(\rho) \partial_x (\delta^2 \partial_x^2 \rho + 2\theta)) + \alpha\rho e^\theta - \alpha\rho^2 e^{-\theta}/\rho_0 \quad \text{(SM.43)}$$

then connects the conjugate momentum $\theta$ to the field $\rho$. Note that the LLN equation (i.e. the noise free mean field description) can be obtained from this equation by setting $\theta = 0$:

$$\partial_t \rho = \partial_x (\mathcal{D}_e \partial_x \rho - \delta^2 \rho D(\rho) \partial_x^3 \rho) + \alpha\rho - \alpha\rho^2/\rho_0 \quad \text{(SM.44)}$$



\end{document}